 \documentclass[preprint2]{aastex}

\def\dsm{$\mathrm{M}_\odot$}
\def\dsr{$\mathrm{R}_\odot$}

\shorttitle{Solar models}
\shortauthors{Wuming Yang}

\begin{document}


\title{Solar models with new low-metal abundances}

\author{Wuming Yang }
\affil{Department of Astronomy, Beijing Normal University,Beijing 100875, China}
\email{ yangwuming@bnu.edu.cn; yangwuming@ynao.ac.cn}


\begin{abstract}
In the last decade, the photospheric abundances of the Sun had been
revised several times by many observers. The standard solar models
(SSM) constructed with the new low-metal abundances disagree with helioseismic
results and detected neutrino fluxes. The solar model problem has been puzzled 
some stellar physicists for more than ten years. Rotation, enhanced diffusion, 
convection overshoot, and magnetic fields are used to reconcile the new 
abundances with helioseismology. The \textbf{too} low-helium \textbf{subsurface
abundance} in enhanced diffusion models can be improved by the mixing caused 
by rotation and magnetic fields. The problem of the depth of the convective zone
in rotating models can be resolved by convection overshoot. 
Consequently the Asplund-Grevesse-Sauval rotation model including 
overshooting (AGSR) reproduces the seismically inferred sound-speed and density
profiles, and the convection zone depth as well as the Grevesse and Sauval (GS98) 
model computed before. But this model fails to reproduce the surface helium abundance
which is 0.2393 ($2.6$ $\sigma$ away from the seismic value) and neutrino fluxes.
The magnetic model called AGSM keeps the agreement of the AGSR and improves 
the prediction of the surface helium abundance. The observed separation
ratios $r_{02}$ and $r_{13}$ are reasonably reproduced by AGSM. 
Moreover, neutrino fluxes calculated by this model are not far from
the detected neutrino fluxes and the predictions of previous works.
\end{abstract}


\section{Introduction}
\subsection{The constraints of helioseismology}
Since \cite{lod03} and \cite{asp04, asp05} revised the ratio of 
heavy-element abundance to hydrogen abundance of the Sun 
(Z/X) from the old 0.023 \citep[hereafter GS98]{gre98} to 0.0177 \citep{lod03}
or 0.0165 \citep[hereafter AGS05]{asp05}, the solar model problem (or 
the solar abundance problem) that standard solar models (SSM) constructed 
with the AGS05 mixtures disagree with the seismically inferred 
sound-speed and density profiles, convection zone (CZ) depth, 
and CZ helium abundance has been perplexed some solar physicists, see 
\cite{tur04}. The seismically inferred CZ base radius is $0.713\pm0.003$ 
\dsr{} \citep{chr91} or $0.713\pm0.001$ \dsr{} \citep{bas97},
and CZ helium abundance is $0.2485\pm0.0035$ \citep{bas04, ser10}.

Many models have been proposed to resolve the problem. 
It was found that a 11\%$-$20\% increase in OPAL opacities 
at the base of the CZ (BCZ) can reconcile the low-Z models with
helioseismology \citep{bas04, mon04, bah04b}. 
However, \cite{bad05} and \cite{guz05} showed that the
increase in the opacities is no more than about 3.0\% near the BCZ.
\cite{ant05} and \cite{bah05} found that an increase in neon 
abundance along with small increases in the other abundances could
solve the problem with AGS05 models. However, \cite{sch05} and \cite{you05}
showed that the Ne/O ratio is indeed consistent with the value given by AGS05.
\cite{asp04} suggested that increased diffusion and settling of helium and 
heavy elements might be able to resolve these disagreements. Several groups, 
e.g., \cite{bas04}, \cite{mon04}, \cite{guz05}, and \cite{yang07} considered
the effects of enhanced diffusion. They found that enhanced diffusion depletes
the CZ helium abundance to well below the seismically inferred value and
leaves the position of the BCZ too shallow. 

Recently, \cite{lod09}, \cite{asp09}, and \cite{caf10} reevaluated the 
spectroscopic abundances of the Sun. The heavy-element abundance was revised
to $Z/X=0.0181$ and $Z=0.0134$ \citep[hereafter AGS09]{asp09}, $Z/X=0.0191$ 
and $Z=0.0141$ \citep{lod09}, and $Z/X=0.0211$ and $Z=0.0154$ \citep{caf10}.
Compared to the SSMs with GS98 mixtures, solar models constructed in 
accordance to the AGS09 mixtures also disagree with the seismically inferred 
sound-speed and density profiles, CZ depth, and CZ helium abundance
\citep{ser09, ser10, ser11, guz10, bi11, tur10, tur11, lop13, lop14, lep15}. 
The conclusion has been that it is difficuilt to match simultaneously the new 
$Z/X$ and helioseismic constraints for sound-speed and density profiles, 
CZ depth, and CZ helium abundance. A resolution to the solar model problem 
remains elusive.

The hypothesis of a large error in the new photospheric abundance estimate
has now been ruled out \citep{lod09, asp09, caf10}. Assuming accretion of metal-poor
gas at the beginning of the main sequence of the Sun, \cite{cas07}, \cite{guz10},
and \cite{ser11} found that the sound-speed profile of accretion model matches very
well that of the GS98 SSM below $R/R_{\odot}=0.5$, but the bump below the CZ remains
quite prominent. Furthermore, the CZ depth of the accretion model is too shallow.
Moreover, \cite{ser11} considered accretion of metal-rich gas. Metal-rich accretion
can bring the depth of the CZ into agreement with the seismic value, but that the 
resulting surface He abundance in such models is too low.

The effects of a convective overshoot were considered by \cite{mon06}, 
\cite{cas07}, and \cite{guz10}. The overshoot below the CZ allows \cite{mon06}
and \cite{cas07} to recover the good CZ radius but can not improve 
sound-speed and surface He abundance \citep{mon06, cas07, guz10}. 
Moreover, the CZ radius of overshoot model of \cite{guz10} is not in
agreement with the seismic value; and the effects of a convective overshoot
does not inhibit He diffusion \citep{guz10}. \cite{guz10} also considered 
the effects of mass loss. They found that the sound-speed agreement 
is considerably improved by including early mass loss. 
But the CZ depth is still too low.

\cite{tur10} constructed
dynamical solar models including a detailed transport of angular momentum
and chemicals due to internal rotation that includes meridional circulation
and shear-induced turbulence. They found that the impact of the rotation
on the solar structure is rather small, and the sound speed is only very
slightly modified when internal rotation is introduced with respect to
the measured internal rotation profile. Their work sustains the idea that 
the Sun was not at the beginning a rapid rotator, and that other dynamical
processes should be included to better reproduce the observed solar profile
and to better describe the young active Sun. Furthermore, \cite{tur11} 
concluded that about $20\%$ of the present discrepancy could come from 
the incorrect description of the early phase of the Sun, its activity,
its initial mass, and mass-loss history. \cite{tur11} also found that
the solar initial mass could have been larger by $20\%-30\%$ than
the present solar mass, and that the Sun could have transformed 
about $2.5\%-4\%$ of the energy produced during the early evolutionary 
stages into other form of energy through kinetic and magnetic energies. 
\cite{tur11} put forward an important view that a transformation 
of nuclear energy into kinetic and magnetic energies during the solar life
must be considered. \cite{zhang14} found that the density profile below the 
CZ is sensitive to the turbulent kinetic flux, which supports 
the view of \cite{tur11}.

Moreover, \cite{lep15} constructed a standard solar model with the
new OPAS opacity tables. Their results show that OPAS opacities improve 
the sound-speed profile but the bump below the CZ remains quite 
prominent (see their Figures 3 and 4).

\subsection{The neutrino flux constraints}
In the last decade, important progress has been made
in the \textbf{detection of neutrinos} \citep{ahm04, 
bel11, bel12} and in the \textbf{prediction of solar neutrino
fluxes} \citep{tur01, tur04, tur10, tur11a, tur12, bah01, bah04a,
cou03, lop13, lop14}. The Sudbury Neutrino Observatory
confirmed the existence of solar neutrino oscillations \citep{ahm04}.
The $^{8}$B, $^{7}$Be, $pp$, and $pep$ neutrino fluxes were determined
\citep{ahm04, bel11, bel12}. 

Seismic observations constrain mainly the external layers 
of the \textbf{Sun and the internal seismic sound speed,} but neutrino 
fluxes probe the real center of the Sun. Neutrinos have complemented
helioseismology in diagnosing the structure of the Sun \citep{tur11, tur12}. 
The boron neutrino flux is strongly dependent on solar central 
temperature \citep{cou03, tur10, tur12}. The $pep$ neutrino flux
is dramatically dependent on the luminosity of the Sun. Therefore,
it represents a powerful probe of the physics of the nuclear 
region of the Sun, in parallel to the information introduced 
in the seismic model \citep{tur12, lop13}. Thus neutrinos
provide a direct constraint on the solar core, see the 
review of \cite{tur11a}.

The solar neutrinos have been used to diagnose the temperature
profile in the Sun's core, measure the radial electronic 
density of matter of the Sun \citep{tur12, lop13} and the 
strength of magnetic field in the radiative region \citep{cou03}, 
and look for standing g-modes of the Sun \citep{lop14}.

\subsection{The uncertainties on the elemental diffusion and mixing}

The rates of element diffusion are enhanced by applying a straight multiplier
to the diffusion velocity, as has been done by \cite{bas04}, \cite{mon04},
\cite{guz05}, and \cite{yang07}. The theoretical error of the gravitational 
settling rate is of the order of about 15\% \citep{tho94}.
Our multipliers of the diffusion coefficients are very high, 
despite the fact that there is no obvious physical justification
for such high multipliers, as has been pointed out by \cite{bas04}
and \cite{guz05}. 

Gravitational settling reduces the surface helium abundance
by roughly $10\%$ ($\approx 0.03$ by mass fraction) below its initial
value but this estimate is still subjected to great uncertainties \citep{pro91}.
\cite{tho94} find the same order of magnitude but state that 
the effects of meridional circulation and of turbulent mixing are ignored
in their models. \cite{pro91} estimated
the effect of turbulent mixing on gravitational settling. They found 
that turbulent mixing could partly inhibit the settling of helium
by about $40\%$. Their turbulent diffusion coefficient is very different
from those given by \cite{pin89} and \cite{zahn97}. \cite{pro91} concluded
that it is unlikely that turbulent mixing of any kind can reduce the 
amount of surface He settling in solar models by more than a factor 
of $2$ below that calculated for pure diffusion models. 
\cite{tur10} \textbf{finds that rotational mixing has a limited effect
on the diffusion for realistic internal rotation rate.}

The observed subsurface mass fraction of helium is 
$0.2485\pm0.035$ \citep{bas04}. As the initial helium 
abundance of the Sun varies between $0.273\pm0.006$ and $0.278\pm0.006$ 
\citep{ser10}, the roughly $10\%$ reduction in the surface helium 
abundance is generally required in many best solar models, 
such as BP00 \citep{bah01}, BP04 \citep{bah04a}, and SSeM \citep{cou03}, 
to reproduce seismic results. So, in order to keep such 
surface helium abundance, the rate of element diffusion must 
be enhanced in our rotating models. \cite{guz05}, \cite{bas08}, 
\cite{tur11} and \cite{lep15} also pointed out that there may
be an insufficient treatment of the microscopic diffusion.

In order to restore agreement between seismic constraints and models 
that are constructed with the AGS09 abundances, based on
the rotating models of \cite{pin89} and \cite{yang07}, 
the combinations of diffusion, rotation, convection overshoot, 
and magnetic fields are considered in this work.

The effects of rotation on the structure and evolution of stars include 
centrifugal effect and rotational mixing \citep{pin89, yang06, tur10,
yang13a, yang13b}. The mechanisms of rotational mixing, considered in this work, 
include the hydrodynamical instabilities \citep{pin89} and secular shear
instability \citep{zahn97}. 

The overshoot of convection extends the region of chemical mixing by a distance 
$\delta_{ov}H_{p}$ below the BCZ that is determined by Schwarzchild 
criterion, where $H_{p}$ is the local pressure scaleheight and $\delta_{ov}$ 
is a free parameter. The full mixing of chemical compositions is assumed in the
overshoot region in our models. The paper is organized as follow: 
the properties of the different solar models are presented in section 2, 
results are given in section 3, a discussion and summary in section 4.

\section{Proposed Solar Models}

The solar models are computed by using Yale Rotation Evolution Code 
\citep{pin89, gue92, yang07} in its rotation and non-rotation configurations.
The OPAL equation-of-state (EOS2005) tables \citep{rog02} and OPAL opacity 
tables \citep{igl96} were used, supplemented by the \cite{fer05} opacity tables
at low temperature. The opacity tables were reconstructed using GS98 or 
AGS09 mixtures. The low-temperature opacity tables with AGS09 mixtures 
were computed by Ferguson for \cite{ser09}. Convection is treated according to 
the standard mixing-length theory. The diffusion and settling of both helium and 
heavy elements are computed by using the diffusion coefficients of \cite{tho94}.
All models are calibrated to the present solar radius $6.9598\times10^{10}$ cm,
luminosity $3.844\times10^{33}$ erg $\mathrm{s}^{-1}$, mass $1.9891\times10^{10}$ g,
and age $4.57$ Gyr. The initial $X_{init}$ and $Z_{init}$, and mixing-length 
parameter $\alpha$ are adjusted to match the constraints of luminosity and radius
within about $10^{-4}$. The values of the parameters are summarized in Table \ref{tab1}.

The nuclear reaction rates were evaluated with the subroutine of \cite{bah92},
updated by \cite{bah95} and \cite{bah01} using the reaction data in \cite{ade98}, 
\cite{gru98}, and \cite{mar00}. 

Following \cite{li01}, we use the toroidal ($B_{t}$) and poloidal ($B_{p}$)
components to express a magnetic filed vector $\mathbf{B}=(B_{t}, B_{p})$. We
assume $|B_{t}|\gg|B_{p}|$ in the radiative region. The magnetic energy density
variable $\chi$ is defined as 
\begin{equation}
 \chi = (B^{2}/8\pi)/\rho,
\end{equation}
where $B=(B_{t}^{2}+B_{p}^{2})^{1/2}$ is the magnitude of the magnetic field vector.
The magnetic pressure $P_{\chi}$ can be written as 
\begin{equation}
 P_{\chi}=\chi\rho.
\end{equation}
The total pressure $P_{T}$ is written as
\begin{equation}
  P_{T}=P+P_{\chi}.
\end{equation}
The equation of state becomes
\begin{equation}
  \frac{d\rho}{\rho}=\alpha\frac{dP_{T}}{P_{T}}-\delta\frac{dT}{T}
  -\psi\frac{d\chi}{\chi},
\end{equation}
where
\begin{equation}
  \alpha=(\frac{\partial\ln\rho}{\partial\ln P_{T}})_{T,\chi},
  \delta=-(\frac{\partial\ln\rho}{\partial\ln T})_{P_{T},\chi},
  \psi=-(\frac{\partial\ln\rho}{\partial\ln \chi})_{P_{T},T}.
\end{equation}
The energy conservation equation becomes
\begin{equation}
 \frac{\partial L}{\partial M_{r}}=\epsilon-T\frac{dS_{T}}{dt},
\end{equation}
where
\begin{equation}
 TdS_{T}=dU+PdV+d\chi
\end{equation}
is the first law of thermodynamics including magnetic fields \citep{li01}.
The total internal energy $U_{T}=U+\chi$ and the total entropy $S_{T}=S+\chi/T$.

The transport process of angular momentum and chemical compositions
caused by magnetic fields is treated as a diffusion process, i.e.,
    \begin{equation}
       \frac{\partial \Omega}{\partial t}=f_{\Omega}
       \frac{1}{\rho r^{4}}\frac{\partial}{\partial r}(\rho r^{4}D_{m}
       \frac{\partial \Omega}{\partial r}) \,,
      \label{diffu1}
    \end{equation}
    \begin{equation}
        \frac{\partial X_{i}}{\partial t}=f_{c}f_{\Omega}\frac{1}{\rho r^{2}}
       \frac{\partial}{\partial r}(\rho r^{2}D_{m}\frac{\partial X_{i}}
        {\partial r})\,,
      \label{diffu2}
    \end{equation}
where $D_{m}=r^{2}\Omega B_{p}^{2}/B^{2}$ is the diffusion coefficient,
$f_{\Omega}$ and $f_{c}$ are a constant between 0 and 1 \citep{yang06, yang08}, 
respectively. The strength and spatiotemporal distribution of magnetic
fields inside a star are poorly known. \cite{yang06} assumed that
$B_{p}^{2}/B^{2}$ is a constant. Due to $|B_{t}|\gg|B_{p}|$, here we
take $|B|\approx|B_{t}|$. The magnetic field compositions $|B_{t}|$ 
and $|B_{p}|$ in the radiative region are calculated by 
using equations (22) and (23) of \cite{spr02}. The values 
of $f_{\Omega}$ and $f_{c}$ are shown in Table \ref{tab1}. 

Solid body rotation is assumed in the CZ. And the distribution 
of the magnitude of magnetic fields is assumed to be a Gaussian profile 
\citep{li01} in the CZ, i.e.,
\begin{equation}
 B = B_{BCZ}\exp[-\frac{1}{2}(r-r_{BCZ})^{2}/\sigma^{2}],
\end{equation}
where the $B_{BCZ}$ is determined by the equations of \cite{spr02}.
The magnitude of $B_{BCZ}$ is of the order of about 
$2\times10^{3}$ Gauss. The value of the $\sigma$ is 
equal to $0.05$. With these assumptions, the magnitude of magnetic
fields is about $1$ Gauss at $M_{r}/M_{\odot}=0.9998$.  
Our model is a simplified representation of the convective magnetic
field. Our simple description does not respect the stability and 
specific configuration of magnetic fields. \cite{duez09, duez10}
studied the impact of large-scale magnetic fields on the structure
and evolution of stars. They pointed out that a mixed poloidal-toroidal
configuration is needed for the fields to survive over evolution
timescales.

We constructed the following five models: 1) GS98M, a standard solar model constructed
by using GS98 mixture opacities; 2) AGS1, a standard model with AGS09 mixture opacities;
3) AGS2, an enhanced diffusion model with AGS09 mixture opacities; 4) AGSR, 
same as AGS2 but with rotation \citep{pin89, yang07} and convection overshoot; 
5) AGSM, same as AGSR but including the effects of magnetic fields.
The initial rotation velocity $V_{init}$ of models is a free parameter 
and is adjusted to obtain the surface rotation velocity of about $2.0$ 
km $\mathrm{s}^{-1}$ at the age of 4.57 Gyr. The values of the parameters
of diffusion and convection overshoot are shown in Table \ref{tab1}. 

\section{Calculation Results}
\subsection{Standard and enhanced diffusion models}
\begin{figure}
\includegraphics[angle=-90, scale=0.7]{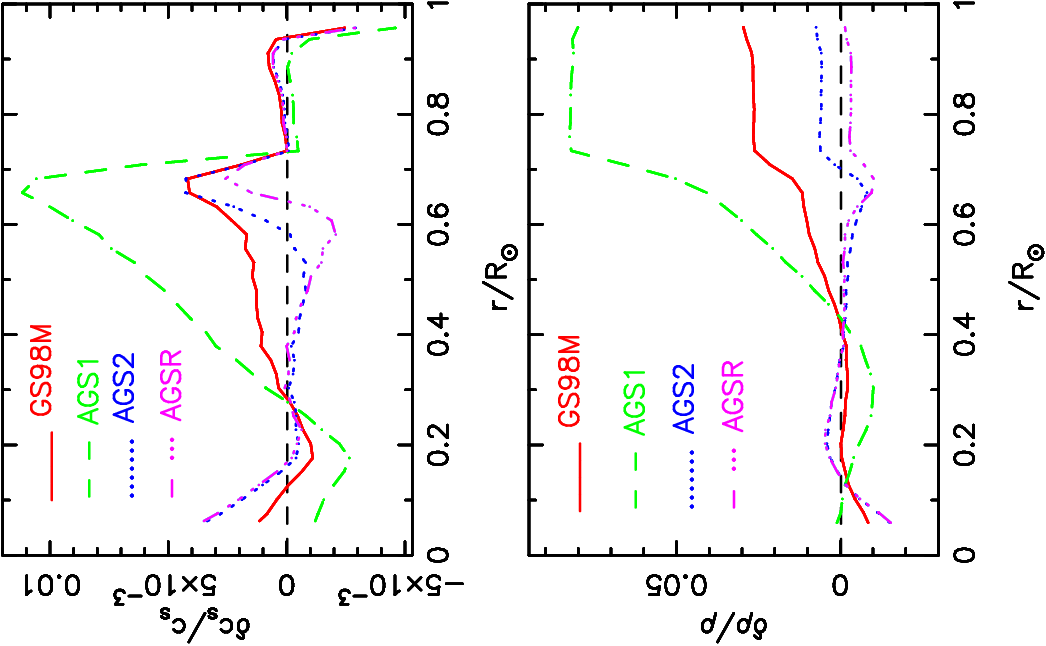}
\caption{The relative sound-speed difference and density difference,
in the sense (Sun-Model)/Model, between solar models and helioseismological
results. The helioseismological sound speed and density are given in \cite{bas09}.
\label{fig1}}
\end{figure}

\begin{figure}
\includegraphics[angle=-90, scale=0.7]{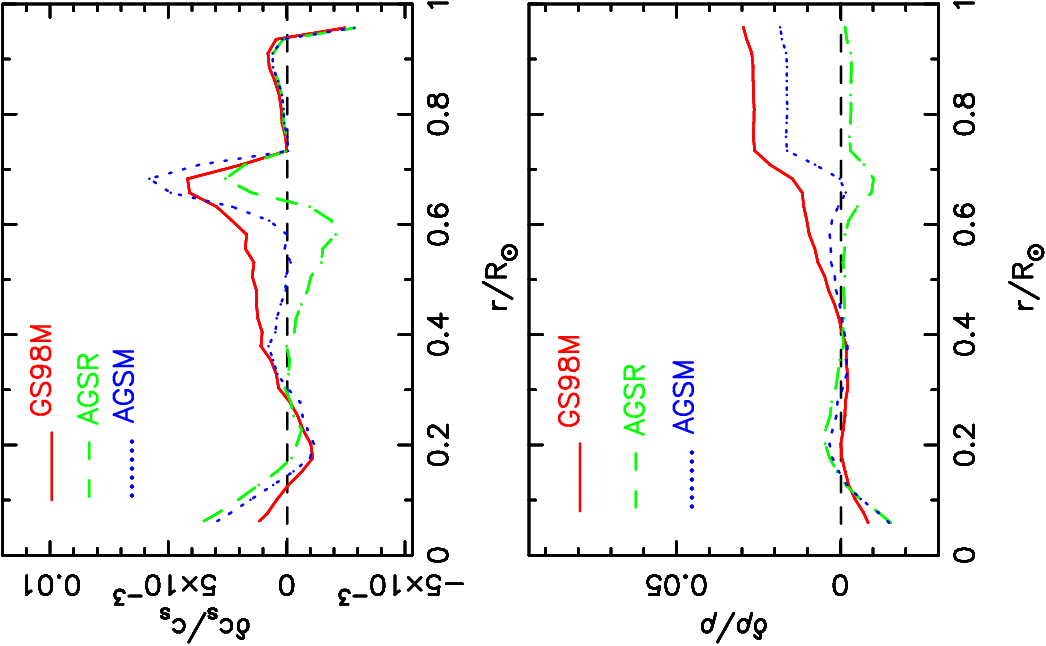}
\caption{The relative sound-speed difference and density difference,
in the sense (Sun-Model)/Model, between solar models and helioseismological
results. The helioseismological sound speed and density are given in \cite{bas09}.
\label{fig2}}
\end{figure}

The sound speed and density of our ad hoc models are compared to those inferred
in \cite{bas09} using Birmingham Solar-Oscillations Network (BiSON) \citep{cha96}
data. The position of the CZ base, the surface helium abundance and heavy-element
abundance of the models are listed in Table \ref{tab1}. Compared to the seismically
inferred CZ base radius \citep{bas97} and helium abundance \citep{bas04}, 
the position of the CZ base of AGS1 is too shallow and its surface
helium abundance is too low. Figures \ref{fig1} and \ref{fig2} show 
the relative differences between the calculated and inferred sound speed 
and density profiles. The relative sound-speed and density differences 
of AGS1 are too large compared to those of GS98M, as has been shown
by many authors \citep{ser09, ser10, ser11, guz10, bi11, tur10, tur11, lep15}.

\begin{figure}
\includegraphics[angle=-90, scale=0.7]{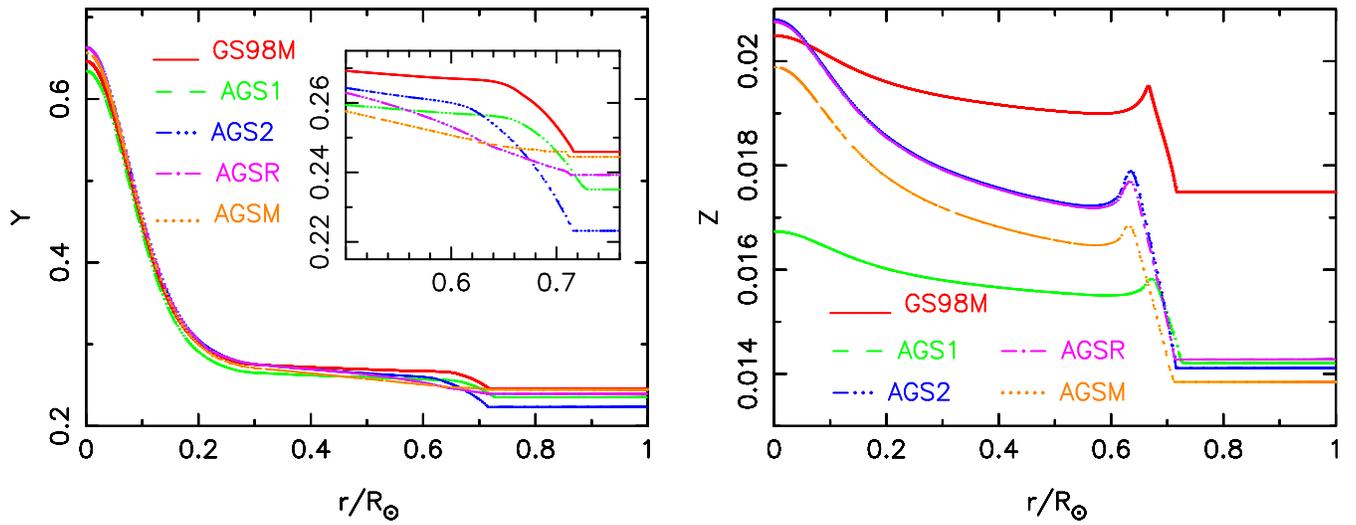}
\caption{Radial distributions of helium and heavy-element mass fraction
of different models. 
\label{pgzx}}
\end{figure}

In order to obtain the model that can restore agreement with helioseismology, 
we constructed the models with an enhanced diffusion rate, following
\cite{bas04}, \cite{guz05}, and \cite{yang07}. However, we have no physical
justification for these multipliers that are required to restore agreement 
with helioseismology in our calculations. Due to the effect of the enhanced 
diffusion, radial distributions of element abundances of AGS2 are
closer to those of GS98M in the radiative region than those of AGS1
(see Figure \ref{pgzx}). Thus the CZ base radius, the sound-speed and
density profiles of AGS2 are close to those of GS98M. 
However, the surface helium abundance of 0.223 is 6 $\sigma$
away from the seismically inferred value \citep{bas04}.

\subsection{Rotating model}
Rotational and turbulent mixing can reduce the surface helium
settling \citep{pro91, yang06, tur10, yang13a, yang13b}. But the position
of the CZ base of rotating models with low Z is too shallow \citep{yang07}.
\cite{mon06} and \cite{cas07} showed that convective overshoot can bring 
the depth of the CZ into agreement with the seismic value. Therefore,
in order to resolve the low-helium problem and the CZ depth problem, 
we constructed rotating solar model AGSR with a convection overshoot
that is described by parameter $\delta_{ov}$. The convection
overshoot brings the CZ base radius into agreement with the seismic value.
By full mixing the material in the overshoot region with that in the 
convective envelope, convection overshoot should lead to an increase in the 
surface helium abundance. The mass of overshoot region for $\delta_{ov}=0.1$
or $0.2$ is about $0.1\%-0.2\%$ \dsm{}, but the mass of the CZ is 
around $2.5\%$ \dsm{}. Moreover, the gradient of helium abundance at 
the base of the convective envelope, caused by microscopic diffusion 
and element settling, has been partly flattened by rotational mixing.
Thus the increase in helium abundance caused by the overshooting 
is very small. Compared with the effect of rotational mixing, 
the effect of the overshooting on the surface helium abundance 
is negligible. The convection overshoot does not improve the surface
helium abundance.

Due to the fact that rotational velocity is low, centrifugal
effect is negligible in AGSR. The impact of rotation on the solar
model mainly derives from the effects of rotational mixing that
rely on hydrodynamical instabilities considered in the model. 
The initial element abundances of AGSR are almost the same as 
those of AGS2. Figure \ref{pgzx} shows that rotational mixing 
considered in AGSR do not affect clearly the distributions 
of element abundances below $0.5$ \dsr{}. Thus the models AGS2 
and AGSR have almost the same central temperature, density, element
abundances (see Table \ref{tab1}), and neutrino fluxes (see Table
\ref{tab2}). However, the rotational mixing below the CZ base partly 
counteracts the surface He and heavy-element settling. Therefore,
the He and heavy-element abundances of AGSR are higher than those
of AGS2 above $0.65$ \dsr{} but are lower than those of AGS2
between about $0.5$ and $0.65$ \dsr{}. So the density
of AGSR is larger than that of AGS2 above $0.65$ \dsr{} but 
is smaller than that of AGS2 between about $0.5$ and $0.65$ \dsr{}.
As a consequence, the sound-speed profile of AGSR is significantly
changed between about $0.5$ and $0.7$ \dsr{}.

Rotational mixing in AGSR reduces the amount of surface He settling
by about $29\%$ and the amount of surface heavy element settling by
about $5\%$. Although the rate of element diffusion is multiplied 
by a factor of $2$ for helium abundance, the surface helium 
abundance is reduced by only about $14\%$ below its initial value in AGSR. 

The CZ base radius of $0.714$ \dsr{} of AGSR is consistent with 
the seismically inferred value of $0.713\pm0.001$ \dsr{} \citep{bas97}. 
The absolute values of relative sound-speed difference, $\delta c/c$,
and density difference, $\delta\rho/\rho$, between AGSR and the Sun
are less than 0.0035 and 0.016, respectively. These values are slightly 
less than those of GS98M. The density profile of AGSR is better than 
that of GS98M. The surface helium abundance of 0.23928 of AGSR is in 
agreement with the seismically inferred value at the level of $2.6$ $\sigma$.

\begin{figure}
\includegraphics[angle=-90, scale=0.7]{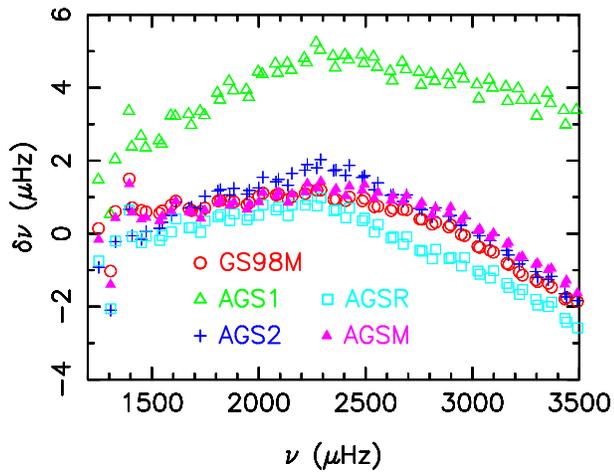}
\caption{Observed minus theoretical frequency vs theoretical frequency
of different models for low-degree modes. The frequencies
of low-degree p-modes of the Sun are observed by GOLF \& VIRGO \citep{gar11}. 
\label{pgdf}}
\end{figure}

However, Figures \ref{pgdf} shows that the frequencies of 
low-degree p-modes of AGSR are not as good as those of GS98M.
Neutrino fluxes calculated from AGSR disagree with those predicted
by \cite{cou03}, \cite{tur04}, and \cite{bah04a}.

\subsection{Rotating model including the effects of magnetic fields}
\cite{tur10, tur11} suggested that magnetic fields should be 
considered in solar models. The effects of magnetic pressure, 
magnetic energy, and the mixing of angular momentum and chemical 
compositions caused by magnetic fields are considered in AGSM.

Compared with the effect of rotational and magnetic mixing,
the effect of the overshooting on the surface helium abundance
is negligible in AGSM. But the convection overshoot allows
us to recover the CZ base radius at the level of $1\sigma$. 
The surface He abundance is $0.2445$, which agrees with seismic 
value at the level of $1.1 \sigma$. The surface Z/X ratio of this
model is $0.0187$. The mixing caused by rotation and magnetic fields
efficiently reduces the surface He settling. Thus the He abundance
of AGSM is higher than that of AGSR above $0.63$ \dsr{} but is lower
than that of AGSR below $0.63$ \dsr{}. The mixing reduces 
the amount of surface He settling by about $47\%$ in AGSM, which is 
slightly higher than $40\%$ that was obtained by \cite{pro91} 
using different diffusion coefficient but does not exceed the upper
limit of $50\%$ assumed by \cite{pro91}. The surface He abundance
is reduced by only $11\%$ ($\approx0.03$ by mass fraction) below 
its initial value. Although the rate of element diffusion is multiplied 
by a factor of $2$ for helium abundance, the $11\%$ reduction in
the surface He abundance is consistent with that of GS98M. 

The relative sound-speed difference and density difference
between AGSM and the Sun, $\delta c/c$ and $\delta\rho/\rho$, 
are less than 0.0058 and 0.019, respectively. The density profile 
of AGSM is as good as that of GS98M (see Figure \ref{fig2}). 
The bump of sound-speed profile mainly appears between 0.63 and 
0.71 \dsr{}. Between 0.1 and 0.6 \dsr{}, the sound speed and density
of this model match very well with those inferred by \cite{bas09}. 

Figure \ref{pgdf} represents the differences between observed 
frequencies of low-degree p-modes \citep{gar11} and those calculated 
from different models, which shows that the agreement between the observed
and theoretical frequencies is improved by the effects of magnetic fields. 

\begin{figure}
\includegraphics[angle=-90, scale=0.7]{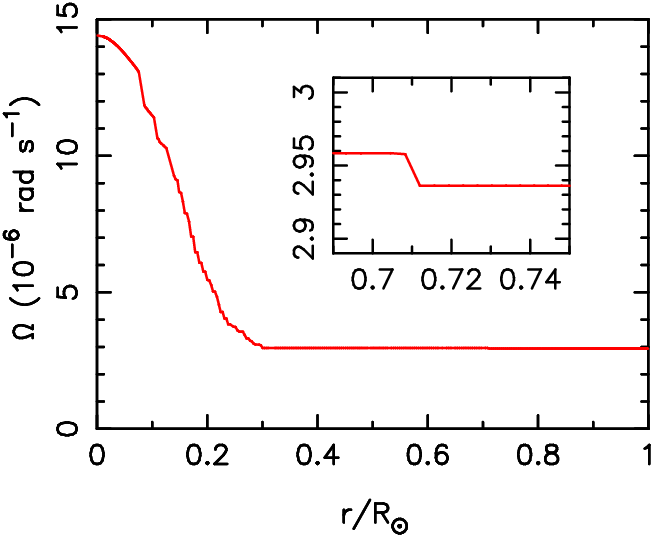}
\includegraphics[angle=-90, scale=0.7]{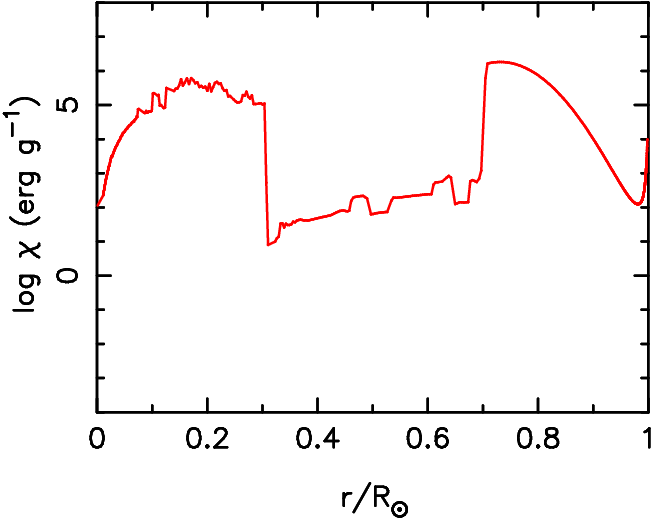}
\caption{Radial distributions of angular velocity and magnetic 
energy density of AGSM. 
\label{pgoe}}
\end{figure}

The initial rotation velocity of AGSR and AGSM is 
$5.8$ km s$^{-1}$, which is consistent with the conclusion
that solar rotation velocity is in the range of $5-10$ km s$^{-1}$
at zero-age main sequence \citep{tur10}. The surface velocity
of both AGSR and AGSM is about $2$ km s$^{-1}$. Efficient 
transport of angular momentum flattens the angular velocity 
profile of AGSM between about 0.3 and 0.7 \dsr{} (see
Figure \ref{pgoe}), which is consistent with seismically 
inferred result \citep{cha99a} and is compatible with the 
first observation of gravity modes (see figures 2 and 4 of \cite{gar07}
or figure 8 of \cite{tur10}). The central angular velocity
is $5$ times as large as the surface angular velocity. 
The total angular momentum of AGSM is $2.40\times10^{48}$ 
g cm$^{2}$ s$^{-1}$ that is closer to $1.94\pm0.05\times10^{48}$ 
g cm$^{2}$ s$^{-1}$ inferred by \cite{kom03} than 
$3.85\times10^{48}$ g cm$^{2}$ s$^{-1}$ of AGSR. 

The magnetic energy is mainly stored in the radiative region
(see Figure \ref{pgoe}). \cite{cou03} showed that an internal large-scale
magnetic field cannot exceed a maximum strength of about $3\times10^{7}$ G
in the radiative region. The strength of magnetic field of AGSM is 
less than about $3\times10^{4}$ G in the radiative region but is less
than around $3\times10^{3}$ G in the CZ.

\begin{figure}
\includegraphics[angle=-90, scale=0.7]{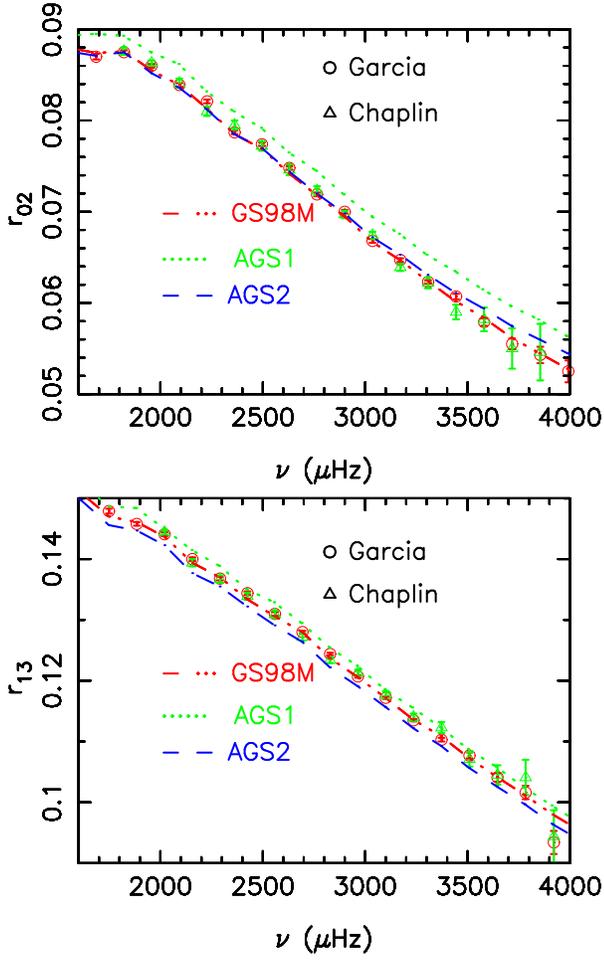}
\caption{The distributions of the ratios of small to large separations,
$r_{02}$ and $r_{13}$, as a function of frequency. The circles show 
the ratios calculated from the frequencies observed by GOLF \& VIRGO \citep{gar11}, 
while the triangles represent the ratios computed from the frequencies observed by 
BiSON \citep{cha99b}.
\label{pgd1}}
\end{figure}

\begin{figure}
\includegraphics[angle=-90, scale=0.7]{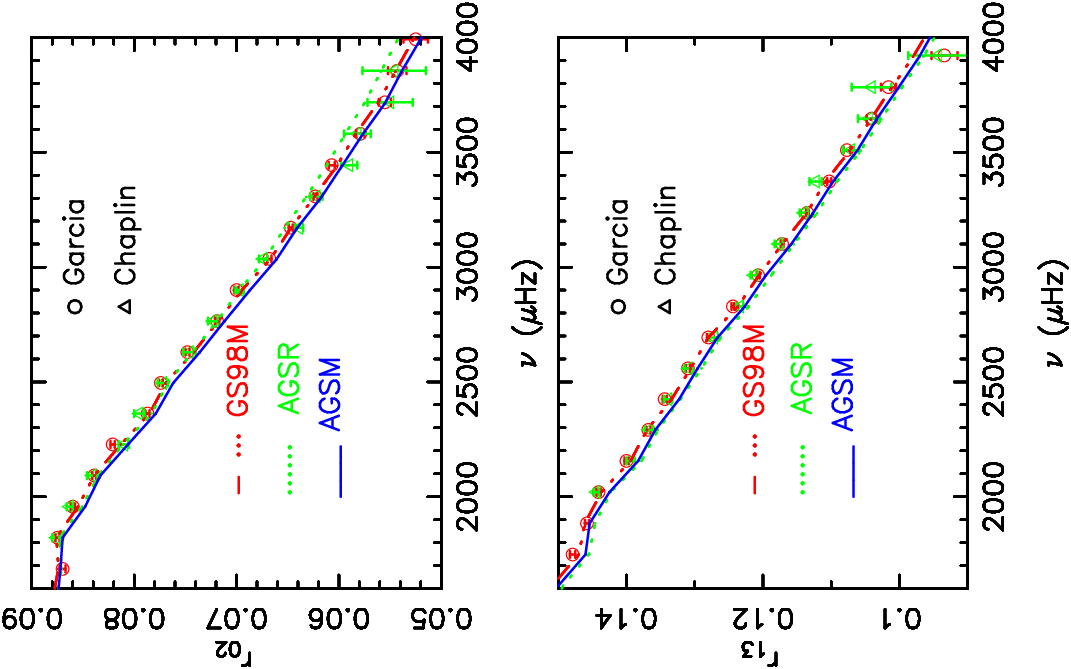}
\caption{The distributions of ratios $r_{02}$ and $r_{13}$
as a function of frequency. The circles show the ratios calculated
from the frequencies observed by GOLF \& VIRGO \citep{gar11}, 
while the triangles represent the ratios
computed from the frequencies observed by BiSON \citep{cha99b}.
\label{pgd2}}
\end{figure}

Furthermore, Figures \ref{pgd1} and \ref{pgd2} show the distributions of 
ratios of small to large separations, $r_{02}$ and $r_{13}$, of these models 
as a function of frequency. The ratios are essentially independent of 
the structure of outer layer and are only determined by the interior
structure of stars \citep{rox03}. The distributions of the ratios of AGS1 
disagree with those calculated from the observed frequencies of \cite{cha99b}
or \cite{gar11}. This indicates that interior structures of AGS1 do not 
match those of the Sun. Although the GS98M model performs much better
than other models, the observed ratios are almost reproduced by AGSR 
and AGSM. 

\subsection{Prediction of neutrino fluxes }
Neutrino fluxes can provide a strict constraint on the core of 
the Sun, which is independent on helioseismology \citep{tur11a}. 
Table \ref{tab2} lists detected and predicted solar neutrino fluxes.
The BP04 and SSeM are the best model of \cite{bah04a}, 
\cite{cou03} and \textbf{\cite{tur11}}, respectively. 
These models not only are in agreement with helioseismic results,
but can reproduce the measured neutrino fluxes \citep{bah04a,
tur01, tur04, tur10, tur11, cou03, tur11a}. We compare neutrino
fluxes computed from our models with those predicted by BP04 and SSeM.

The neutrino fluxes calculated from GS98M are in good 
agreement with those predicted by \cite{bah01} except $^{7}$Be
neutrino flux. The $^{7}$Be neutrino flux computed from GS98M is
also larger than those predicted by BP04 and SSeM. The neutrino 
fluxes calculated from AGS1, AGS2, and AGSR are obviously 
different from those predicted by GS98M, BP04, and SSeM. However,
the fluxes of $pp$, $pep$, $^{8}$B, $^{13}$N, and $^{15}$O neutrinos
of AGSM are almost in agreement with those of BP04 \citep{bah04a} and 
SSeM \citep{tur01, tur04, tur11, cou03, tur11a}. The $^{17}$F neutrino 
flux of AGSM is also in agreement with that of BP04 but is higher than 
that of SSeM. The relative difference between the neutrino fluxes of 
AGSM and those of BP04 is less than $2.0\%$.

The $hep$ neutrino flux of AGSM is consistent with those of GS98M 
but is higher than that of BP04. The discrepancy between
$hep$ neutrino flux of AGSM and that of BP04 can be attributed to 
the nuclear cross section factor $S_{0}(hep)$ which only affects
the flux of $hep$ neutrinos and does not affect the calculated 
fluxes of other neutrinos \citep{bah01}. Comparing the neutrino
fluxes calculated from AGS1, AGS2, AGSR, and AGSM with those 
predicted by BP04 and SSeM and the detected neutrino fluxes,
one can find that the neutrino-flux agreement is considerably 
improved by the effects of magnetic fields and that AGSM is 
obviously better than \textbf{pure SSM} low-Z models. 

\section{Discussion and Summary}
The equatorial velocity of about $2$ km s$^{-1}$, i.e. 
$2.94\times10^{-6}$ rad s$^{-1}$, is taken as the surface 
velocity of models at the age of the Sun. \textbf{This value}
is slightly higher than $2.72\times10^{-6}$ rad s$^{-1}$ 
adopted by \cite{kom03}. Thus the total angular momentum of AGSM is
slightly larger than that inferred by \cite{kom03}. The low angular
velocity can be achieved by a slow initial rotation. The sound speed
profile is practically unchanged when a slower initial rotation is
adopted, which is consistent with the result of \cite{tur10}.

The rates of element diffusion are enhanced by multipling 
a factor of 2 to the diffusion velocity of He and a factor of 2.5
to the diffusion velocity of heavy elements. The effect of mixing
caused by rotation and magnetic fields reduces the amount of surface
He settling by about $47\%$. So the surface He abundance is reduced
by only $11\%$ below its initial value. A roughly $10\%$ reduction 
in the surface He abundance has been proved by many best solar models.
Solar activity and helioseismology show the limitation of SSM and call
for the dynamical solar model including dynamical processes 
\citep{tur10, tur11a}. In order to keep a roughly $10\%$ reduction in
the surface He abundance, the rates of element diffusion 
must be enhanced in our dynamical solar models.

The SSM with AGS09 mixtures does not agree with seismic constraints
for sound-speed and density profiles, CZ base radius, and CZ helium 
abundance. Thus we calculated the dynamical solar models that include
the effects of enhanced diffusion, rotation, convection overshoot,
and magnetic fields. The discrepancies between models with AGS09 
mixtures and helioseismic results can be significantly reduced
by the enhanced diffusion. However, the surface helium abundance 
of the enhanced diffusion model is too low. The surface
He settling can be partially counteracted by rotational mixing.
Thus the low-helium problem can \textbf{be} resolved by the effect of rotation
to a great extent. Convection overshoot aids in resolving the problem
of shallow CZ base position but does not affect clearly the 
surface helium abundance. Thus the CZ base radius of models with
convection overshoot are consistent with the seismically inferred value. 
This reflects that rotation and convection overshoot may be important
in the evolution of the Sun. Moreover, rotation plays an important 
role in the formation of the extended main-sequence turnoff of
intermediate-age massive star clusters \citep{yang13a, yang13b}; 
and convection overshoot plays an essential role in explaining some of 
characteristics of solar-like oscillations of stars \citep{yang15}. 
These indicate that the effects of rotation and convection overshoot 
should \textbf{not be} ignored in the evolutions of stars.

Although the rotating AGSR model can reproduce the seismically
inferred sound-speed and density profiles and the CZ base radius, 
the neutrino fluxes calculated from this model are not in agreement
with those predicted by BP04 \citep{bah04a} and SSeM 
\citep{cou03, tur04, tur11a}. Moreover, the total angular momentum
of $3.85\times10^{48}$ g cm$^{2}$ s$^{-1}$ of AGSR is too high.
\cite{tur11} suggested that 
the effects of magnetic field must be considered during the solar
life. We consider the effects of magnetic pressure, magnetic energy,
and magnetic mixing. The surface He abundance and neutrino fluxes
are significantly improved by the effects of magnetic field.

In this work, we constructed dynamical solar models with the AGS09
mixtures in which the effects of enhanced diffusion, rotation, convection
overshoot, and magnetic fields were included. We obtained two models: 
AGSR and AGSM that almost restore agreement with helioseismology.
Compared to GS98 SSM, AGSR can reproduce the CZ base radius, 
the sound-speed and density profiles. The surface helium abundance 
of about 0.2393 of this model is $2.6$ $\sigma$ away from the 
seismically inferred value of \cite{bas04}. The position of the CZ base 
of AGSM agrees with the seismically inferred value at the level of 
$1$ $\sigma$. The surface helium abundance of 0.2445 of AGSM
agrees with the seismic value at the level of $1.1\sigma$.
The density profile of this model is as good as that of GS98M.
The sound-speed profile of AGSM matches very well that inferred in \cite{bas09}
between 0.1 and 0.6 \dsr{}. The bump of the sound speed mainly appears
between 0.63 and 0.71 \dsr{}. The relative difference of the sound speed
between AGSM and the Sun, $\delta c/c$, is less than 0.0058 in the tacholine.
Moreover, the observed separation ratios $r_{02}$ and $r_{13}$ are almost
reproduced by AGSM; the initial helium abundance of AGSM is in agreement
with $0.273\pm0.006$ given by \cite{ser10} at the level of $1$ $\sigma$. 

The $^{8}$B neutrino flux predicted by AGSM 
agrees with the detected one and the prediction of SSeM
at the level of about $1\sigma$. The fluxes of $pp$, $pep$, 
and $^{7}$Be neutrinos computed from AGSM are not far 
from the detected ones and the predictions of BP04 and SSeM.
The improvement in the sound-speed and density profiles and
the CZ radius mainly derives from the effects of the enhanced 
diffusion and overshooting which are dependent of the dynamical 
processes, while the improvement in the surface helium abundance 
and neutrino fluxes basically comes from the effects of the
dynamical processes. Although the GS98 SSM performs much better
than the low-Z models as a whole, the agreement between the 
low-Z models and seismic and neutrino results is improved 
by the effects of the dynamical processes.

\acknowledgments The author thanks the anonymous referee for helpful 
comments which help the author improve this work, and the support from 
the NSFC 11273012, 11273007, 11503039, and the Fundamental 
Research Funds for the Central Universities.

 \clearpage

\clearpage

\clearpage

\begin{table*}
\begin{center}
\caption{Model parameters. \label{tab1}}
\begin{tabular}{cccccc}

\tableline\tableline
 Parameter               & GS98M    & AGS1     & AGS2    &   AGSR   & AGSM \\
\tableline 
$X_{init}$               & 0.70418  & 0.71772  & 0.70334 & 0.7034 &  0.7085 \\
$Y_{init}$               & 0.27638  & 0.266397 & 0.27844 & 0.27841 &  0.274054 \\
$Z_{init}$               & 0.01944  & 0.015883 & 0.01822 & 0.01819 &  0.017446 \\           
$\alpha$                 & 2.213    & 2.1769   & 2.402  &  2.25748 &  2.336 \\                      
$\delta_{ov}$            &    0     &   0      &   0    &   0.1    &  0.2 \\
$V_{init}$ (km s$^{-1}$) &    0     &   0      &   0    &   5.8    &  5.8   \\ 
Multiplier  & 1.0\tablenotemark{a}(1.0)\tablenotemark{b}& 1.0 (1.0) &  2.0 (2.5) & 2.0 (2.5) & 2.0 (2.5) \\
$f_{\Omega}$            &    0     &   0      &   0    &   0      &  0.01  \\
$f_{c}$                 &    0     &   0      &   0    &   0  &  1.0$\times10^{-4}$  \\
\tableline
$T_{c}$ ($10^{6}$ K)     & 15.78    & 15.65    & 16.00   & 15.99   &  15.91 \\
$\rho_{c}$ (g cm$^{-3}$) & 154.64   & 152.14   & 156.39  & 156.38  &  156.01 \\
$X_{c}$                  & 0.3337   & 0.3492   & 0.3162  & 0.3167  &  0.3233 \\
$Y_{c}$                  & 0.6458   & 0.6341   & 0.6630  & 0.6626  &  0.6568 \\
$R_{cz}$/$R_{\odot}$ & 0.716\tablenotemark{c} & 0.729 & 0.716 & 0.714 & 0.712 \\
$Y_{s}$                  & 0.24591  & 0.23507  & 0.22317 & 0.23928 &  0.2445 \\
$Z_{s}$                  & 0.01748  & 0.01420  & 0.01411 & 0.01427 &  0.01385 \\
$(Z/X)_{s}$              & 0.0237   & 0.0189   & 0.0185  & 0.0191  &  0.0187 \\
$V_{e}$ (km s$^{-1}$)    &   0      &    0     &    0    & 1.90    &  2.04 \\
$J_{tot}\times10^{48}$ (g cm$^{2}$ s$^{-1}$) & 0 & 0 & 0 & 3.85 &  2.40  \\
\tableline
\end{tabular}
\tablenotetext{a}{ The multiplier for the diffusion coefficient of
the helium;} \tablenotetext{b}{ The multiplier for the diffusion
coefficient of the heavy elements;} \tablenotetext{c}{ Using OPAL
EOS96, \citet{bah04b} obtained $R_{cz}$ = 0.7155 $R_{\odot}$.}
\end{center}
\end{table*}

\begin{table*}
\begin{center}
\caption{Predicted solar neutrino fluxes from models. 
The table shows the predicted fluxes, in units of $10^{10}(pp)$,
$10^{9}(^{7}\mathrm{Be})$, $10^{8}(pep,^{13}\mathrm{N}, ^{15}\mathrm{O})$,
$10^{6}(^{8}\mathrm{B}, ^{17}\mathrm{F})$, and $10^{3}(hep)$ 
$\mathrm{cm}^{2} \mathrm{s}^{-1}$. The BP04 is the best model of
\cite{bah04a} and has the GS98 mixtures. The SSeM is the best 
standard model that reproduces the seismic sound speed \citep{cou03, tur11a}.
The old SSeM has high-metal abundances \citep{cou03},
but the new SSeM has the low-metal abundances \citep{tur04, tur11a}.
\label{tab2}}
\begin{tabular}{cccccccccc}

\tableline\tableline
Source    & GS98M & AGS1 & AGS2 & AGSR&  AGSM &  BP04 & old SSeM & new SSeM & Measured\\
\tableline 
 $pp$     & 5.95 & 6.01 & 5.88 & 5.88 &  5.91 &  5.94 & 5.92     & .... & 6.06$^{+0.02}_{-0.06}$\tablenotemark{(a)} \\
 $pep$    & 1.40 & 1.43 & 1.38 & 1.38 &  1.39 &  1.40 & 1.39     & .... & 1.6$\pm0.3$\tablenotemark{(b)}\\
 $hep$    & 9.47 & 9.77 & 9.24 & 9.22 &  9.33 &  7.88 & ....     & .... & .... \\ 
 $^{7}$Be & 5.11 & 4.72 & 5.60 & 5.57 &  5.33 &  4.86 & 4.85     & 4.72 & $4.84\pm0.24$\tablenotemark{(a)}\\
 $^{8}$B  & 5.22 & 4.43 & 6.48 & 6.41 &  5.82 &  5.79 & 4.98     & 5.31$\pm$0.6 & $5.21\pm0.27\pm0.38$\tablenotemark{(c)}\\
 $^{13}$N & 5.46 & 3.91 & 6.47 & 6.42 &  5.71 &  5.71 & 5.77     & .... & .... \\
 $^{15}$O & 4.83 & 3.39 & 5.88 & 5.82 &  5.13 &  5.03 & 4.97     & .... & .... \\
 $^{17}$F & 5.59 & 3.89 & 6.88 & 6.82 &  5.99 &  5.91 & 3.08     & .... & .... \\
\tableline
\end{tabular}
\tablenotetext{(a)}{\cite{bel11}.}
\tablenotetext{(b)}{\cite{bel12}.}
\tablenotetext{(c)}{\cite{ahm04}.}
\end{center}
\end{table*}

\end{document}